\documentclass[aps,superscriptaddress,nofootinbib,onecolumn,a4paper,11pt]{revtex4-2}
\usepackage{amsfonts,amsmath,amssymb}
\usepackage[english]{babel}
\usepackage{graphicx}
\usepackage{natbib}
\usepackage{bm}
\usepackage{dcolumn}
\usepackage{graphicx}
\usepackage{color}
\usepackage{slashed}
\usepackage[usenames,dvipsnames]{xcolor}
\usepackage{soul}
\usepackage{verbatim}
\usepackage{ulem}
\usepackage{float}
\usepackage{mathtools}

\definecolor{forestgreen}{rgb}{0.13, 0.55, 0.13}
\definecolor{internationalorange}{rgb}{1.0, 0.31, 0.0}

\def\gmatrix{{\cal G}}

\def\trmin{{\rm tr}}
\def\mf{{\mbox{\tiny MF}}}

\DeclareMathOperator\arctanh{arctanh}

\DeclareMathOperator\arcsinh{arcsinh}

\begin{document}

\vspace*{1cm}

\title{\Large $\pi^0$-$\gamma$ mixing in the presence of a strong magnetic field}
\author{D.~Gomez~Dumm}
\email{dumm@fisica.unlp.edu.ar}
\affiliation{Instituto de F\'isica La Plata, CONICET $-$ Departamento de F\'isica, Facultad de Ciencias Exactas,
Universidad Nacional de La PLata, C.C. 67, (1900) La Plata, Argentina}
\affiliation{CONICET, Rivadavia 1917, (1033) Buenos Aires, Argentina}
\author{S.~Noguera}
\affiliation{Departamento de F\'isica Te\'orica and IFIC, Centro Mixto Universidad
de Valencia-CSIC, E-46100 Burjassot (Valencia), Spain}
\author{N.N.~Scoccola}
\affiliation{CONICET, Rivadavia 1917, (1033) Buenos Aires, Argentina}
\affiliation{Physics Department, Comisi\'{o}n Nacional de Energ\'{\i}a
At\'{o}mica, Avenida del Libertador 8250, 1429 Buenos Aires,
Argentina}

%%%%%%%%%%%%%%%%%%%%%%%%%%%%%%%%%%%%%%%%%%%%%%%%%%%%%%%%%%%%%%%%%%%%%%%%%%%%%%%%%%%%%%%%
\begin{abstract}
We analyze the mixing between the $\pi^0$ meson and the photon in the
presence of a strong uniform magnetic field, in the context of a two-flavor
Nambu-Jona-Lasinio model. It is shown that only one specific photon
polarization state can get mixed with the $\pi^0$ state, a fact that can be
understood on general grounds. For magnetic fields with values up to
1~GeV$^2/e$, it is found that the effect of the mixing on the pion mass and
pion-quark couplings is relatively small (below a level of 15\%), which is
at variance with previous results reported in the literature.
\end{abstract}

%%%%%%%%%%%%%%%%%%%%%%%%%%%%%%%%%%%%%%%%%%%%%%%%%%%%%%%%%%%%%%%%%%%%%%%%%%%%%%%%%%%%%%%%
%\pacs{xx }

\maketitle

\newpage

%%%%%%%%%%%%%%%%%%%%%%%%%%%%%%%%%%%%%%%%%%%%%%%%%%%%%%%%%%%%%%%%%%%%%%%%%%%%%%%%%%%%%%%%
\section{Introduction}

The response of strongly interacting matter to the presence of extremely
large magnetic fields has attracted considerable attention over the past few
decades~\cite{Kharzeev:2012ph,Andersen:2014xxa,Miransky:2015ava,Adhikari:2024bfa}.
To a large extent, this interest is driven by the important role played by
these fields in the physics of the early
Universe~\cite{Vachaspati:1991nm,Grasso:2000wj}, in noncentral high-energy
heavy ion collisions~\cite{Skokov:2009qp,Voronyuk:2011jd} and in the
properties of compact stellar objects such as
magnetars~\cite{Duncan:1992hi,Kouveliotou:1998ze}. In addition, strong
magnetic fields have been shown to induce several remarkable phenomena in
quantum chromodynamics (QCD), including the chiral magnetic
effect~\cite{Kharzeev:2007jp,Fukushima:2008xe,Kharzeev:2015znc}, the
enhancement of the quark-antiquark condensate (magnetic
catalysis)~\cite{Gusynin:1995nb} and the reduction of critical temperatures
for chiral restoration and deconfinement transitions (inverse magnetic
catalysis)~\cite{Bali:2011qj}.

Given this context, the understanding of how the properties of light hadrons
get modified in the presence of an intense magnetic field becomes a highly
relevant task. This is, however, a nontrivial problem, since first-principle
calculations require to deal with QCD in the low-energy, nonperturbative
regime. Consequently, theoretical analyses have been carried out using a
variety of approaches, including Lattice QCD, Chiral Perturbation Theory
and effective quark models for strong interactions (see
Ref.~\cite{Adhikari:2024bfa} and references therein).

An interesting aspect to consider is the mixing between hadronic states with
different spin and/or isospin. Although these mixed states are typically
forbidden by isospin and angular momentum conservation, they can exist in
the presence of an external magnetic field. Effects of this kind have been
recently studied by the authors of the present work within the framework of
the Nambu-Jona-Lasinio (NJL) model. In particular, in
Refs.~\cite{Carlomagno:2022inu,Coppola:2023mmq,Carlomagno:2022arc} it was
shown that magnetic-field-induced mixing between pseudoscalar and
vector/axial-vector channels tends to improve the agreement between model
predictions and available Lattice QCD results.

In this work we focus on another type of mixing induced by the external
magnetic field, namely the $\pi^0$-$\gamma$ mixing. The mixing between
photons and low-mass bosons has been extensively studied in the past,
particularly in the context of scalar particles such as axions (see
e.g.~Ref.~\cite{Raffelt:1987im}). Within strong interaction physics,
magnetic-field-induced $\pi^0$-$\gamma$ mixing was previously analyzed in
Ref.~\cite{Brauner:2017uiu} using leading order Chiral Perturbation Theory.
In that study, non-anomalous one-loop corrections to the pion mass were
neglected. Under that approximations, it was found that the mixing with the
photon induces a significant increase of the neutral pion mass with the
magnetic field.

Motivated by these results, the aim of the present work is to reexamine the
impact of $\pi^0$-$\gamma$ mixing on the neutral pion mass within the NJL
model. In this framework, mesons are described as quantum fluctuations in
the random phase approximation
(RPA)~\cite{Vogl:1991qt,Klevansky:1992qe,Hatsuda:1994pi}, i.e., they arise
from the resummation of an infinite series of quark loop diagrams. In the
presence of a magnetic field, these loops must be evaluated using magnetized
quark propagators. In the terminology of Ref.~\cite{Brauner:2017uiu}, this
corresponds to the inclusion of the non-anomalous contribution to the
$\pi^0$ mass. Consistently, the $\pi^0$-$\gamma$ mixing is also generated by
loops of magnetized quarks. Therefore, this approach has the important
advantage that both contributions to the $\pi^0$ mass are treated on an
equal footing.

Regarding the model setup, special care must be taken in the
regularization of ultraviolet divergences, since the presence of
an external magnetic field can lead to spurious effects. In this
work we adopt the magnetic-field-independent regularization
(MFIR) scheme~\cite{Ebert:1999ht,Menezes:2008qt,Allen:2015paa},
which has proven to be particularly suitable for magnetized
systems~\cite{Avancini:2019wed}.

The article is organized as follows. In Sec.~II.A we introduce the main
theoretical framework and obtain the quadratic terms of the effective action
that are relevant for the $\pi^0$-$\gamma$ mixing. The quantities of
interest (mixing matrix, pion mass, pion wave function renormalization and
pion-quark couplings) are analytically calculated in Sec.~II.B, II.C and
II.D. In Sec.~III we present the corresponding numerical results, while in
Sec.~IV we summarize our work and state our main conclusions. We also
include an appendix in which we collect some mathematical expressions used
for the calculation of polarization functions.

%%%%%%%%%%%%%%%%%%%%%%%%%%%%%%%%%%%%%%%%%%%%%%%%%%%%%%%%%%%%%%%%%%%%%%%%%%%%%%%%%%%%%%%%

\section{Theoretical framework}

\subsection{Bosonized effective action in the NJL model}

Let us start by considering a NJL Lagrangian density in which $u$ and $d$
quarks are coupled to an electromagnetic field ${\cal A}_\mu$. One has
\begin{equation}
{\cal L} \ = \ \bar{\psi}\left(i\,\rlap/\!D-m_{c}\right)\psi
+ \, g \left[ \left(\bar{\psi} \psi\right)^2 + \left(\bar{\psi}\, i \gamma_5 \vec \tau \psi\right)^2 \right]
- \frac{1}{4}\, F_{\mu\nu} F^{\mu\nu}\ ,
\end{equation}
where $\psi=(u\ d)^{T}$, $\tau_i$ are the usual Pauli matrices, and $m_{c}$
is the current quark mass, which is assumed to be equal for $u$ and $d$
quarks. The interaction between the fermions and the electromagnetic field
is driven by the covariant derivative
\begin{equation}
D_{\mu}\ =\ \partial_{\mu}+i\,\hat{Q}\,\mathcal{A}_{\mu}\ ,
\label{covdev}
\end{equation}
where $\hat{Q}=\mbox{diag}(Q_{u},Q_{d})$, with $Q_{u}=2e/3$ and
$Q_{d}=-e/3$, $e$ being the proton electric charge. As usual, $F^{\mu\nu}$
is the electromagnetic field strength tensor, given by $F^{\mu\nu} =
\partial^\mu {\cal A}^{\nu} - \partial^\nu {\cal A}^{\mu}$.
%A summary of
%the notation and conventions used throughout this work can be found
%in App.~\ref{conventions}.

Since we consider the presence of an external magnetic field, we write the
gauge field $\mathcal{A}^{\mu}$ as the sum of an external (classical) field
$\mathcal{A}^{\mu}_{cl}$ and a dynamical electromagnetic field $a^{\mu}$,
i.e., $\mathcal{A}^{\mu} =\mathcal{A}^{\mu}_{cl} + a^{\mu}$. As stated, we
consider here the particular case in which we have an external homogenous
and static magnetic field $\vec{B}$. Without loss of generality, it can be
taken to be orientated along the axis 3, or $z$. Notice that to write down
the explicit form of $\mathcal{A}^{\mu}_{cl}$ one has to choose a specific
gauge. However, it will be seen that our calculations turn out to be
manifestly gauge independent.

Since we are interested in studying the $\pi^0$-$\gamma$ mixing, it is
convenient to bosonize the fermionic theory, introducing scalar and
pseudoscalar fields $\sigma(x)$ and $\vec \pi(x)$, respectively. After
integrating out the fermion fields one gets a bosonized action given by
\begin{eqnarray}
S_{\mathrm{bos}} & = & -i\ln\det\!\big(i\mathcal{D}\big)-\frac{1}{4g}\int d^{4}x\
\Big[\sigma(x)\,\sigma(x)+\vec{\pi}(x)\cdot\vec{\pi}(x)\Big] + S_{\rm em}\ ,
\label{sbos}
\end{eqnarray}
with
\begin{equation}
i\mathcal{D}_{x,x'}\ = \ \big[i\,\rlap/\!D_x-m_{0}- \sigma(x)-i\,\gamma_{5}\,\vec \tau \cdot \vec \pi(x)\big]
\, \delta^{(4)}(x-x') \ ,
\label{dxx}
\end{equation}
where a direct product to an identity matrix in color space is understood.
The pure electromagnetic field contribution $S_{\rm em}$ is given by
\begin{eqnarray}
S_{\rm em} & = & - \frac{1}{4} \int d^{4}x \ F_{\mu\nu} F^{\mu\nu} \nonumber \\
& = & -\frac{V^{(4)}\, B^2}{2} + \frac{1}{2} \int d^{4}x \ a^\mu(x)\, \Big[
\eta_{\mu\nu}\, \partial_\alpha \partial^\alpha - \partial_{\mu}
\partial_{\nu} \Big]\, a^\nu(x)\ ,
\end{eqnarray}
where the first term of the last expression corresponds to the classical
piece and the second one includes the propagating degrees of freedom. Since
we consider fixed values of the external magnetic field, in what follows the
classical contribution to $S_{\rm em}$ will be dropped.

We proceed by expanding the bosonized action in powers of the fluctuations
of the bosonic fields around the corresponding mean field (MF) values. We
assume that the fields $\sigma$ have nontrivial translational invariant MF
values $\bar{\sigma}$, while vacuum expectation values of the pion fields
are zero. Thus, we write
\begin{equation}
\mathcal{D}_{x,x'}\ =\ \mathcal{D}_{x,x'}^{\mbox{\tiny MF}}+\delta\mathcal{D}_{x,x'}\ .
\label{dxxp}
\end{equation}
The MF piece is diagonal in flavor space and includes the interaction of the
underlying quark fields with the classical electromagnetic field
$\mathcal{A}^{\mu}_{cl}$. One has
\begin{equation}
\mathcal{D}_{x,x'}^{\mf}\ =\ {\rm diag}\big(\mathcal{D}_{x,x'}^{\mf,\,u}\,,\,\mathcal{D}_{x,x'}^{\mf,\,d}\big)\ ,
\end{equation}
where
\begin{equation}
\mathcal{D}_{x,x'}^{\mf,\,f}\ =\ -i\,
\left(i\rlap/\partial_x-Q_{f}\; \rlap/\!\mathcal{A}_{cl}(x) - M\right) \ \delta^{(4)}(x-x') \ ,
\end{equation}
with $f=u,d$. Here $M=m_{c}+\bar{\sigma}$ stands for a quark effective
mass. On the other hand, the correction $\delta\mathcal{D}_{x,x'}$ in
Eq.~(\ref{dxxp}) reads
\begin{equation}
\delta\mathcal{D}_{x,x'} \ =\ i \Big[ \hat Q \ \rlap/\!a(x) + \delta \sigma(x)+i\gamma_{5}\,\vec \tau \cdot
\delta \vec \pi(x) \Big] \, \delta^{(4)}(x-x')\ .
\label{dxx}
\end{equation}

The MF action per unit volume is given by
\begin{equation}
\frac{S_{\mathrm{bos}}^{\mbox{\tiny MF}}}{V^{(4)}}\ =-\ \frac{\bar{\sigma}^{2}}{4g}-\frac{iN_{c}}{V^{(4)}}\sum_{f=u,d}\int d^{4}x\,d^{4}x'
\ \trmin_{D}\,\ln\left(\mathcal{S}_{x,x'}^{\mf,\,f}\right)^{-1}\ ,
\label{seff}
\end{equation}
where $\trmin_{D}$ stands for the trace over Dirac indices, and
$\mathcal{S}_{x,x'}^{\mf,\,f}=\big(i\mathcal{D}_{x,x'}^{\mf,\,f}\big)^{-1}$
is the MF quark propagator in the presence of the magnetic field. Its
explicit expression can be written as
\begin{equation}
\mathcal{S}_{x,y}^{\mf,\,f}=e^{i\Phi_{Q_{f}}(x,y)}\int\frac{d^{4}p}{(2\pi)^{4}}\ e^{-i\,p(x-y)}
\,\bar{S}^{f}(p_{\parallel},p_{\perp})\ ,
\label{uno}
\end{equation}
where $\Phi_{Q}(x,y)$ is the so-called Schwinger phase associated to the
classical field, and $\bar{S}^{f}(p_{\parallel},p_{\perp})$ is given by
\begin{eqnarray}
\bar{S}^{f}(p_{\parallel},p_{\perp}) & = & -\,i\int_{0}^{\infty}d\sigma\ \exp\!\bigg[\!-i\sigma\Big(M^{2}-p_{\parallel}^{2}+\vec{p}_{\perp}^{\ 2}\,\dfrac{\tan(\sigma B_{f})}{\sigma B_{f}}-i\epsilon\Big)\bigg]\nonumber \\
 &  &
\times\left[\left(p_{\parallel}\cdot\gamma_{\parallel}+M\right)(1-s_f\,\gamma^{1}\gamma^{2}
\tan(\sigma B_{f}))-\frac{\vec{p}_{\perp}\cdot\vec{\gamma}_{\perp}}{\cos^{2}(\sigma B_{f})}\right]\ ,
\label{sfp_schw}
\end{eqnarray}
with $B_f = |B Q_f|$ and $s_f={\rm sign}(B Q_f)$. Here we have defined the
``parallel'' and ``perpendicular'' four-vectors
\begin{equation}
p_{\parallel}^{\mu}=(p^{0},0,0,p^{3})\ , \qquad\qquad
p_{\perp}^{\mu}=(0,p^{1},p^{2},0)\ ,
\end{equation}
and equivalent definitions have been used for $\gamma_\parallel$,
$\gamma_\perp$. The Schwinger phase, which is a gauge dependent quantity,
reads
\begin{eqnarray}
\Phi_{Q}(x,y) = Q_f \int_{x}^y d\xi_\mu \left[{\cal A}_{cl}^\mu(\xi) +
\frac{1}{2}\ F_{cl}^{\mu\nu}\ (\xi_\nu-y_\nu)\right]\ .  \label{sp}
\end{eqnarray}
In fact, since we deal here with a neutral pion, Schwinger phases cancel out
in quark loops and will play no effective role in our calculations.

Let us consider the quark-antiquark condensates $\phi_{f}\equiv\langle\bar{\psi}_{f}\psi_{f}\rangle$.
For each flavor $f=u,d$ we have
\begin{eqnarray}
\phi_{f}=iN_{c}\int\frac{d^{4}p}{(2\pi)^{4}}\ \trmin_{D}\,\bar{S}^{f}(p_{\parallel},p_{\perp})\ ,
\end{eqnarray}
which is clearly gauge independent. The integral in this expression is
divergent and has to be properly regularized. As stated in the Introduction,
we use here the magnetic field independent regularization (MFIR) scheme: for
a given unregularized quantity, the corresponding (divergent) $B\to0$ limit
is subtracted and then it is added in a regularized form. Thus, the
quantities can be separated into a (finite) ``$B=0$'' part and a
``magnetic'' piece. Notice that, in general, the ``$B=0$'' part still depends
implicitly on $B$ (e.g.\ through the values of the dressed quark masses
$M$); hence, it should not be confused with the value of the studied
quantity at vanishing external field. To deal with the divergent ``$B=0$''
terms we use here a 3D cutoff regularization. Thus, we obtain
\begin{equation}
\phi_{f}^{{\rm reg}}\ =\ \phi^{0,\,{\rm reg}}\,+\,\phi_{f}^{{\rm mag}}\ ,
\end{equation}
where
\begin{equation}
\phi^{0,{\rm reg}}=-N_{c}\,M\,I_{1}^{\rm reg}\ ,\qquad\qquad
\phi_{f}^{{\rm mag}}=-N_{c}\,M\,I_{1f}^{{\rm mag}}\ ,
\label{phif}
\end{equation}
with
\begin{eqnarray}
I_{1}^{\rm reg} & = & \dfrac{\Lambda^2}{2 \pi^2} \left[ \,\sqrt{1+R_\Lambda^2} +
R_\Lambda^2 \,\ln\left( \dfrac{R_\Lambda}{1 + \sqrt{1 + R_\Lambda^2}}\right) \right] \ ,
\label{I1freg} \\[3mm]
I_{1f}^{{\rm mag}} & = &
\dfrac{B_{f}}{2\pi^{2}}\left[\ln\Gamma(x_{f})-\left(x_{f}-\dfrac{1}{2}\right)
\ln x_{f}+x_{f}-\dfrac{\ln{2\pi}}{2}\right]\ .
\label{i1}
\end{eqnarray}
Here, $\Lambda$ stands for the 3D momentum cutoff, and we have defined
$R_\Lambda=M/\Lambda$ and $x_{f}=M^{2}/(2B_{f})$. The corresponding gap
equation for $M$, obtained from $\partial S_{\mathrm{bos}}^{\mbox{\tiny
MF}}/\partial\bar{\sigma}=0$, can be written as
\begin{equation}
M \ = \ m_{c}- 2g \sum_f \phi_{f}^{{\rm reg}}\ .
\label{gapeqs}
\end{equation}

Next we consider quadratic terms in the expansion of $S_{\rm bos}$ in powers
of $\delta\mathcal{D}_{x,x'}$. We just concentrate in neutral pion
contributions, since charged pions do not mix with the photon. Since
Schwinger phases cancel out in fermion loops, the corresponding polarization
functions turn out to be both gauge and translational invariant. Moreover,
momentum conservation implies that they have to be diagonal in the momentum
basis. After a Fourier transformation the relevant terms of the bosonized
action can be written as
\begin{eqnarray}
S_{\mathrm{bos}}^{{\rm neut}}\
&=&-\ \frac{1}{2}\int \frac{d^{4}q}{(2\pi)^4} \
\bigg[ \, \delta \pi^0(-q) \ {\gmatrix}_{\pi^0\pi^0}(q)\ \delta \pi^0(q)
\ + \ a_\mu(-q) \ {\gmatrix}_{\gamma\gamma}^{\mu\nu}(q)\ a_\nu(q)
\nonumber \\
& & + \ \delta \pi^0(-q) \ {\gmatrix}_{\pi^0\gamma}^{\mu}(q)\ a_\mu(q)
\ + \  a_\mu(-q) \ {\gmatrix}_{\gamma\pi^0}^{\mu}(q)\ \delta \pi^0(q)\,\bigg]\ ,
\label{sbos}
\end{eqnarray}
where
\begin{eqnarray}
{\gmatrix}_{\pi^0\pi^0}(q) &=& \frac{1}{2 g} - \sum_f \Pi^f_{\pi^0\pi^0}(q)\ ,
\label{gpi0pi0} \\
{\gmatrix}_{\gamma\gamma}^{\mu\nu}(q) &=&
 q^2 \eta^{\mu\nu} - q^{\mu} q^{\nu} -
 \sum_f \Pi^{f,\mu\nu}_{\gamma\gamma}(q)\ , \\
{\gmatrix}_{\pi^0\gamma}^{\mu}(q) &=& -\sum_f \Pi^{f,\mu}_{\pi^0\gamma}(q)\ , \\
{\gmatrix}_{\gamma\pi^0}^{\mu}(q) &=& -\sum_f \Pi^{f,\mu}_{\gamma\pi^0}(q)\ .
\end{eqnarray}
In the above expressions, the polarization functions $\Pi^{f}_{PP'}(q)$,
where $P,P'=\pi^0$ or $\gamma$, are given by
\begin{equation}
\Pi^f_{PP'}(q) \ = \ -i N_c\,\int \frac{d^{4}q}{(2\pi)^4}\ \trmin_{D}\!
\left[ \,i \bar{S}^{f}(p_{\parallel}^{+},p_{\perp}^{+}) \ \Gamma_P^f \ i
\bar{S}^{f}(p_{\parallel}^{-},p_{\perp}^{-})\ \Gamma_{P'}^f \right]\ ,
\label{polfun}
\end{equation}
where $\Gamma_{\pi^0}^f = i\kappa_f\gamma_5$, $\Gamma_{\gamma}^{f,\mu}=
Q_f\gamma^\mu$, with $\kappa_u = 1$, $\kappa_d=-1$, and $p^{\pm}=p\pm q/2$.

Let us proceed with the evaluation of the loop integrals in
Eq.~(\ref{polfun}). The expression for the $\pi^0$-$\pi^0$ polarization
function using the MFIR scheme is given e.g.~in
Refs.~\cite{Carlomagno:2022inu,Coppola:2023mmq}. One has
\begin{equation}
\Pi^f_{\pi^0\pi^0}(q) \ = \ C_{\pi\pi}^{0,{\rm reg}}(q^2) \, + \,
C^{f,{\rm mag}}_{\pi\pi}(q_\parallel^2,q_\perp^2)\ ,
\label{pipi0pi0}
\end{equation}
where
\begin{eqnarray}
C_{\pi\pi}^{0,{\rm reg}}(q^2) & = & N_c\,\Big[ I_1^{\rm reg} \, -\, q^2
I_2^{\rm reg}(q^2)\Big] \ , \\
C^{f,{\rm mag}}_{\pi\pi}(q_\parallel^2,q_\perp^2) \ &=&
\ \frac{N_{c}}{8\pi^{2}}\int_{0}^{\infty}dz\,\int_{-1}^{1}d\beta\ \bigg\{
e^{-z\,\omega^f(q_\parallel^2,q_\perp^2)}\;
\bigg[ \left(M_{f}^{2}+\frac{1}{z}+\frac{1-\beta^{2}}{4}\;q_{\parallel}^{2}\right)
\,\frac{B_{f}}{\tanh(zB_{f})}  \nonumber \\
& & \ \ +\,\frac{B_{f}^{2}}{\sinh^{2}(zB_{f})}
\left(1+\frac{\cosh(zB_{f})-\cosh(\beta z B_{f})}{2B_{f}\sinh(z B_{f})}\;
q_{\perp}^2\right) \bigg]
  \nonumber  \\
& & \ \ -\, \frac{ e^{-z\,\omega_0(q^2)}}{z}
\ \left( M^2+\,\frac{2}{z}+\frac{1-\beta^{2}}{4}\;q^{2}\right) \bigg\}\ .
\label{cppfull}
\end{eqnarray}
Here $I_{1}^{\rm reg}$ is given by Eq.~(\ref{I1freg}), and we have used the
definitions
\begin{eqnarray}
I_2^{\rm reg}(q^2) & = & -\,\frac{1}{4\pi^2}\,\left[\arcsinh(1/R_\Lambda)
- \sqrt{1/r_q-1}\,\arctan\left(\frac{1}{\sqrt{(1+R_\Lambda^2)\,(1/r_q-1)}}\right)
\right]\ ,
\nonumber \\
\omega^f(q_\parallel^2,q_\perp^2) \ &=& \ M^{2}-\frac{1-\beta^{2}}{4}\;q_{\parallel}^{2}
-\frac{\cosh(zB_{f})-
\cosh(\beta zB_{f})}{2\,z\,B_{f}\,\sinh(zB_{f})}\;q_{\perp}^2\ ,
\nonumber \\
\omega_0(q^2) &=& M^2 \,-\,\frac{1-\beta^{2}}{4}\,q^{2}\ ,
\end{eqnarray}
with $r_q = q^2/(4M^2)$. It is assumed that the pion four-momentum squared
is below the threshold $q^2=4M^2$, which corresponds to the formation of
on-shell quark pairs.

The mixing contributions to the effective action are driven by the $\pi^0$-$\gamma$ and
$\gamma$-$\pi^0$ polarization functions. These are found to be given by
\begin{equation}
\Pi^{f,\mu}_{\pi^0\gamma}(q) \ = \ - \Pi^{f,\mu}_{\gamma\pi^0}(q) \ = \
-i\,C^f_{\pi\gamma}(q_\parallel^2,q_\perp^2)\, v^\mu\ ,
\end{equation}
where we have introduced a four-vector $v$ defined by $v^\mu =
(q^3,0,0,q^0)$, and the function $C^f_{\pi\gamma}(q_\parallel^2,q_\perp^2)$
is given by
\begin{equation}
C^f_{\pi\gamma}(q_\parallel^2,q_\perp^2) \ = \
\frac{1}{8\pi^{2}}\, N_c\, s_{f} \, M\,
\kappa_f \,Q_f B_{f}\ \int_0^{\infty}dz\,\int_{-1}^{1}d\beta\
e^{-z\,\omega^f(q_\parallel^2,q_\perp^2)}\ .
\label{cpigam}
\end{equation}
This result is consistent with the expressions obtained in
Ref.~\cite{Coppola:2023mmq} for the mixing between $\pi^0$ and $\rho^0$
mesons in a similar framework.

Finally, we consider the $\gamma$-$\gamma$ polarization function. The
expression for $\Pi^{f,\mu\nu}_{\gamma\gamma}$ can be written in terms of
three transverse tensors, namely
\begin{equation}
P_0^{\mu\nu} = q^2\, \eta^{\mu\nu}\,  - q^{\mu} q^{\nu}\ ,  \quad\quad
P_1^{\mu\nu} = q_\parallel^2 \, \eta_{\parallel}^{\mu\nu}\,  - q_{\parallel}^{\mu} q_{\parallel}^{\nu} \ , \quad\quad
P_2^{\mu\nu} =  q_\perp^2 \, \eta_\perp^{\mu\nu}\,  - q_\perp^{\mu} q_\perp^{\nu}\ ,
\end{equation}
where $\eta_\parallel^{\mu\nu}={\rm diag}(1,0,0,-1)$,
$\eta_\perp^{\mu\nu}={\rm diag}(0,-1,-1,0)$,
$\eta^{\mu\nu}=\eta_\parallel^{\mu\nu}+\eta_\perp^{\mu\nu}$. One has
\begin{equation}
\Pi^{f,\mu\nu}_{\gamma\gamma}(q) \ = \ C_{\gamma\gamma,0}^{f,{\rm reg}}(q_\parallel^2,q_\perp^2)
\, P_0^{\mu\nu} \, + \,
C_{\gamma\gamma,1}^f(q_\parallel^2,q_\perp^2) \, P_1^{\mu\nu} \, + \,
C_{\gamma\gamma,2}^f(q_\parallel^2,q_\perp^2) \, P_2^{\mu\nu} \ ,
\end{equation}
where the regularized function $C_{\gamma\gamma,0}^{f,{\rm reg}}
(q_\parallel^2,q_\perp^2)$ can be written as
\begin{equation}
C_{\gamma\gamma,0}^{f,{\rm reg}}(q_\parallel^2,q_\perp^2) \ = \
C_{\gamma\gamma,0}^{f,{\rm mag}}(q_\parallel^2,q_\perp^2) +
C_{\gamma\gamma,0}^{f,0,{\rm reg}}(q^2)\ .
\label{cmagreg}
\end{equation}
The above functions are given by~\cite{Tsai:1974ap,Urrutia:1977xb}
\begin{eqnarray}
\!\!\!\!C_{\gamma\gamma,i}^f(q_\parallel^2,q_\perp^2) &=& - \frac{N_c Q_f^2}{16 \pi^2} \, B_f
\int_{0}^{\infty}dz\,\int_{-1}^{1}d\beta\ e^{-z\,\omega^f(q_\parallel^2,q_\perp^2)}
\, f_i(z,\beta)\ , \quad \ i=1,2\ , \label{c12} \\
\!\!\!\!C_{\gamma\gamma,0}^{f,{\rm mag}}(q_\parallel^2,q_\perp^2) &=& - \frac{N_c Q_f^2}{16 \pi^2} \,
\int_{0}^{\infty}dz\,\int_{-1}^{1}d\beta
\bigg[ B_f\,e^{-z\,\omega^f(q_\parallel^2,q_\perp^2)}\, f_0(z,\beta) -
e^{-z\,\omega_0(q^2)}\, \frac{(1-\beta^2)}{z}\bigg]\ , \label{c1mag} \\
\!\!\!\!C_{\gamma\gamma,0}^{f,0,{\rm reg}}(q^2) &=& - \frac{N_c Q_f^2}{16 \pi^2} \,
\int_{0}^{\infty}dz\,\int_{-1}^{1}d\beta
\bigg[ e^{-z\,\omega_0(q^2)} - e^{-z\,\omega_0(0)}\bigg] \frac{(1-\beta^2)}{z}\ ,
\label{c1reg}
\end{eqnarray}
with
\begin{eqnarray}
f_0(z,\beta) & = &\frac{ \cosh(\beta z B_{f}) - \beta \coth(z B_{f})
\sinh(\beta z B_{f})}{\sinh(z B_{f})} \ , \nonumber \\
f_1(z,\beta) & = & (1 - \beta^2) \coth(z B_{f})\, - \, f_0(z,\beta)
\ , \nonumber \\[2mm]
f_2(z,\beta) & = & \frac{2\big[\!\cosh( z B_{f}) - \cosh(\beta z B_{f})\big]}{\sinh^3(z B_{f})}
\, - \, f_0(z,\beta)\ .
\end{eqnarray}

Concerning the regularization of the coefficient of the $P_0^{\mu\nu}$ term
($C_{\gamma\gamma,1}^f$ and $C_{\gamma\gamma,2}^f$ are finite and do not
have to be regularized), we have used once again the MFIR scheme, separating
a ``magnetic'' contribution $C_{\gamma\gamma,0}^{f,{\rm mag}}$ and a
``$B=0$'' piece (see Eqs.~(\ref{c1mag}) and (\ref{c1reg})). Then, as shown
in Eq.~(\ref{c1reg}), the (divergent) ``$B=0$'' contribution
$C_{\gamma\gamma,0}^{f,0}(q^2)$ has been regularized by subtracting the
value of this function at $q^2=0$. This prescription has also been used in
previous works, see
e.g.~Refs.~\cite{Tsai:1974ap,Urrutia:1977xb,Hattori:2012je}. The integral
over $z$ in the regularized expression given by Eq.~(\ref{c1reg}) can be
performed analytically, leading to
\begin{equation}
C_{\gamma\gamma,0}^{f,0,{\rm reg}}(q^2) \ = \ \frac{N_c
Q_f^2}{16 \pi^2} \int_{-1}^{1}d\beta\; (1-\beta^{2}) \ln \left[ 1-
(1-\beta^2) \frac{q^2}{4 M^2} \right]\ .
\label{c0reg}
\end{equation}

\subsection{Inverse propagator matrix}

\label{propmatrix}

To determine the $\pi^0$ mass in the presence of the external magnetic field
it is necessary to diagonalize the inverse propagator matrix arising from
the expansion of the bosonized action. Regarding the electromagnetic field
$a_\mu$, we have to consider the two relevant photon states that
contribute to the quadratic action in Eq.~(\ref{sbos}). In the presence of
the external field, the corresponding polarization vectors
$\epsilon^\mu(\lambda)$ are given by (see e.g.\ Ref~\cite{Hattori:2012je})
\begin{equation}
\epsilon^\mu(1) = (0, q^2, -q^1,0)/\sqrt{-q_\perp^2}\ , \qquad\qquad
\epsilon^\mu(2) = (q^3, 0, 0, q^0)/\sqrt{q_\parallel^2}\ ,
\label{polstates}
\end{equation}
where $q$ is the photon four-momentum. These vectors are manifestly
gauge invariant, since they can be written in terms of the electromagnetic
field strength tensor and its dual as
\begin{equation}
\epsilon^\mu(1) = \pm\, F^{\mu\nu}q_\nu /
\sqrt{F^{\sigma\alpha}F^{\sigma\beta}q_\alpha q_\beta}\ ,
\qquad\qquad \epsilon^\mu(2) = \pm\,\tilde F^{\mu\nu}q_\nu /
\sqrt{\tilde F^{\sigma\alpha}\tilde F^{\sigma\beta}q_\alpha q_\beta}\ ,
\end{equation}
where $+$ and $-$ signs correspond to $B>0$ and $B<0$, respectively, and the
dual of $F^{\mu\nu}$ is given by
\begin{equation}
\tilde F^{\mu\nu} \ = \ \frac{1}{2}\,\epsilon^{\mu\nu\alpha\beta}\,F_{\alpha\beta}\ ,
\end{equation}
with the convention $\epsilon^{0123}=1$.

Taking into account the above expressions for the polarization functions, it
is seen that the photon with polarization $\lambda = 1$ decouples, while the
$\lambda = 2$ state becomes mixed with the neutral pion. The inverse
propagator matrix can be written as
\begin{equation}
{\gmatrix} = \left(%
\begin{array}{ccc}
 {\gmatrix}_{\gamma_1\gamma_1}  & 0 &  0 \\
  0 & {\gmatrix}_{\gamma_2\gamma_2}  & {\gmatrix}_{\gamma_2\pi^0} \\
  0 & {\gmatrix}_{\pi^0\gamma_2} & {\gmatrix}_{\pi^0\pi^0}  \\
\end{array}%
\right)\ ,
\end{equation}
where the expression for ${\gmatrix}_{\pi^0\pi^0}$ can be read from
Eqs.~(\ref{gpi0pi0}) and (\ref{pipi0pi0}). The other entries of the matrix
are given by
\begin{eqnarray}
{\gmatrix}_{\gamma_1\gamma_1} & = &
- \Big(1-\sum_f C_{\gamma\gamma,0}^{f,{\rm reg}}\Big)
\,q^2 + \sum_f C_{\gamma\gamma,2}^f\, q_\perp^2 \ , \\
{\gmatrix}_{\gamma_2\gamma_2} & = &
- \Big(1-\sum_f C_{\gamma\gamma,0}^{f,{\rm reg}}\Big)
\,q^2 + \sum_f C_{\gamma\gamma,1}^f\, q_\parallel^2 \ , \label{g2g2} \\
{\gmatrix}_{\gamma_2\pi^0} & = &  -\, {\gmatrix}_{\pi^0\gamma_2} =
i\sum_f C_{\pi\gamma}^{f} \sqrt{q_\parallel^2} \ .
\end{eqnarray}

There is a symmetry reason that explains why the pion just mixes with one
specific photon polarization state. It is important to note that when
studying the $\pi^0$-$\gamma$ mixing in the presence of an external magnetic
field, we are dealing with a massive particle. Therefore, observable
quantities that are meaningless for the photon become fully relevant in this
case. The whole physical system is invariant under rotations around the axis
3, and consequently the third component of total angular momentum,
$J_{3}=\left(\vec{x}\times\vec{q}\right)_{3}+S_{3}$, has to be a good
quantum number. Thus, if we let $\vec{q}_{\perp}=0$, the quantum number
$S_{3}$ will be a good one to characterize the pion states. In
Ref.~\cite{Coppola:2023mmq}, when dealing with $\pi$-$\rho$ meson
mixing in the presence of a magnetic field, it is discussed how $S_{3}$
becomes a good quantum number for the $\rho$ meson. That analysis can also
be applied to the $\pi$-$\gamma$ mixing studied in the present work, showing
that the polarization $\epsilon^{\mu}(2)$ corresponds to a state with
$S_{3}=0$ (hence, it can get mixed with the neutral pion), whereas
$\epsilon^{\mu}(1)$ corresponds to a combination of $S_{3}=+1$ and $S_3=-1$
states.

It is worth noticing that up to now we have just taken into account light
quark contributions to the effective action. However, for the evaluation of
$\gamma$-$\gamma$ terms, all charged fermion loops should be taken into
account. Thus, neglecting the contributions of heavy fermions, the sums over
$f$ in the expressions for ${\gmatrix}_{\gamma_1\gamma_1}$ and
${\gmatrix}_{\gamma_2\gamma_2}$ have to be extended with the inclusion of
electron and muon loops. For these lepton contributions we have to change
$N_c \to 1$ and take electric charges $Q = -e$. In particular, in the case
of the electrons, the loop contribution to ${\gmatrix}_{\gamma_2\gamma_2}$
will include some imaginary part, which can be evaluated by carrying out an
analytical extension of the obtained expressions (see discussion in
App.~\ref{app1}). However, it is found that this absorptive piece is highly
suppressed by a factor $m_e^2/m_\pi^2$, and will be neglected in our
numerical calculations.

\subsection{Neutral pion mass and wave function renormalization}

\label{pimassandwfr}

The $\pi^0$ mass can be obtained by diagonalizing the $\gmatrix$ matrix.
We introduce the states
\begin{eqnarray}
\tilde\pi^0 & = & \,\cos\Theta\; \delta\pi^0 -\, i\, \sin\Theta\;
a^{(2)}\ ,
\nonumber \\
\tilde a^{(2)} & = & \,-\, i\,\sin\Theta\; \delta\pi^0 + \cos\Theta\; a^{(2)}\ ,
\label{deftheta}
\end{eqnarray}
where $a^{(2)}$ is the electromagnetic field state associated to the
polarization vector $\epsilon(2)$. Then we require that the off-diagonal
entries of the rotated matrix ${\tilde\gmatrix}$ vanish. This leads to a
$\Theta$ mixing angle given by
\begin{equation}
\tan 2\Theta \ = \ -\,\frac{ 2i\,\gmatrix_{\pi^0\gamma_2}}
{{\gmatrix}_{\gamma_2\gamma_2}-{\gmatrix}_{\pi^0\pi^0}}\ .
\label{angle}
\end{equation}
Taking into account that Lorentz invariance is broken by the external
magnetic field, we define the $\pi^0$ mass by considering the $\tilde\pi^0$
lowest energy state, which corresponds to the pion rest frame in which $\vec
q_\perp = 0$. The pion mass, given by the condition
\begin{equation}
\tilde\gmatrix_{\tilde\pi^0\tilde\pi^0}\Big|_{(q_\parallel^2,q_\perp^2)=(m_{\pi^0}^2,0)}
 = \ 0\ ,
\end{equation}
can be obtained by solving the equation
\begin{equation}
\Big[\gmatrix_{\pi^0\pi^0} \, \gmatrix_{\gamma_2\gamma_2} -
 |\gmatrix_{\pi^0\gamma_2}|^{\,2}\Big]_{(q_\parallel^2,q_\perp^2)=(m_{\pi^0}^2,0)}
\ = \ 0\ .
\label{eqpimass}
\end{equation}
In general, for $q_\perp^2=0$ the functions $C_{\pi\pi}^{f,{\rm
mag}}$, $C^f_{\pi\gamma}$, $C_{\gamma\gamma,i}^f$ and
$C_{\gamma\gamma,0}^{f,{\rm mag}}$ in Eqs.~(\ref{cppfull}), (\ref{cpigam}),
(\ref{c12}) and (\ref{c1mag}) become simplified. The corresponding
expressions are given in App.~\ref{app1}.

Now, as usual, the pion field wave function has to be renormalized. This is done by
considering the pion mass term in the quadratic action, given by
\begin{eqnarray}
S^{\,\mbox{\tiny quad}}_{\pi^0} &=& -\dfrac{1}{2} \int \frac{d^{4}q}{(2\pi)^4} \
\tilde{\pi}^{0}(-q) \; \tilde {\gmatrix}_{\tilde \pi^0 \tilde \pi^0}(q_\parallel^2,q_\perp^2)\;
\tilde{\pi}^0(q) \ .
\label{actionquadpi0p_2}
\end{eqnarray}
Expanding around the pion pole we have
\begin{eqnarray}
S^{\,\mbox{\tiny quad}}_{\pi^0} &=& \dfrac{1}{2} \int \frac{d^{4}q}{(2\pi)^4} \
\tilde \pi^{0}_R(-q) \, \left(
q_\parallel^2+ u_{\pi^0}^2\, q_\perp^2 -m_{\pi^0}^2 \right) \, \tilde\pi^0_R(q) \ ,
\label{actionquadpi0p_2}
\end{eqnarray}
where the pion field has been renormalized according to $\tilde
\pi^0_R(q)=Z_{\parallel}^{-1/2} \, \tilde \pi^0(q)$, and we have used the
definitions
\begin{equation}
Z_{\parallel,\pi^0}^{-1} \,=\, -\dfrac{\partial\tilde
{\gmatrix}_{\tilde \pi^0 \tilde \pi^0}}{\partial q_\parallel^2}
\Big\rvert_{\!\!{\tiny\begin{array}{l}
               q_\perp^2 = 0 \\
               q_\parallel^2=m_{\pi^0}^2
             \end{array}}} \ ,
\qquad
Z_{\perp,\pi^0}^{-1} \,=\, -\dfrac{\partial\tilde {\gmatrix}_{\tilde \pi^0 \tilde \pi^0}}
{\partial q_\perp^2}\Big\rvert_{\!\!{\tiny\begin{array}{l}
               q_\perp^2 = 0 \\
               q_\parallel^2=m_{\pi^0}^2
             \end{array}}}\ ,
\qquad
u_{\pi^0}^2 = \dfrac{Z_{\parallel,\pi^0}}{Z_{\perp,\pi^0}}\ .
\label{u_pi}
\end{equation}

\subsection{$\tilde\pi^0$-fermion couplings}

From Eq.~(\ref{dxx}) it is seen that one has two contributions to the
effective $\tilde\pi^0$-quark couplings. One of them arises from the
standard vertex $i\gamma_5\kappa_f$, where $\kappa_f$ is associated to the
light quark isospin ($\kappa_u =1$, $\kappa_d=-1$). The second one,
originated from the mixing between the neutral pion and the $a^{(2)}$ field,
is given by $Q_f\gamma^\mu$, where $Q_f$ is the quark electric charge.
Notice that, as discussed in Sec.~\ref{propmatrix} in relation to ${\cal
G}_{\gamma\gamma}$ propagators, this second coupling can be extended to
$Q_{\mathcal F}\gamma^\mu$, being ${\mathcal F}$ any charged fermion (all
charged fermions will couple to the $a^{(2)}$ field). Thus, taking also into
account the pion wave function renormalization, the vertex that couples the
$\tilde\pi_R^0$ to a fermion ${\mathcal F}$ will be given by
\begin{equation}
iG_{\pi^0 q q}\,\gamma_5\,\kappa_{\mathcal F} \,+\,
iG^{\rm em}_{\pi^0}\, \frac{Q_{\mathcal F}}{e}\; \rlap/\!\epsilon_2\ ,
\label{pigvertex}
\end{equation}
where
\begin{equation}
G_{\pi^0 q q} = Z_{\parallel}^{1/2} \cos\Theta\ ,\qquad \qquad
G^{\rm em}_{\pi^0} = G_{\pi^0 q q}\, e\,\tan\Theta\ ,
\label{gp0qq}
\end{equation}
with $\kappa_{\mathcal F}=0$ if ${\mathcal F}\neq u,d$. Notice that,
from the equations in Secs.~\ref{propmatrix} and \ref{pimassandwfr},
$\tan\Theta$ is of order $e$. Hence, as expected, the effective coupling
$G^{\rm em}_{\pi^0}$ is ${\cal O}(e^2)$.

It is clear that, in general, the existence of the $\pi^0$-$\gamma$ mixing
implies that the pion mass state $\tilde\pi_R^0$ will interact with any
charged particle through the coupling between the electromagnetic current
associated to that particle and the $a^{(2)}$ component of the
$\tilde\pi_R^0$ state.

\section{Numerical results}

In this section we discuss the numerical results arising from our
calculation of $\pi^0$ properties. For the model parameters we take the
values $m_0 = 5.419$~MeV, $\Lambda = 6.395$~MeV and $G\Lambda^2=2.136$,
which ---for vanishing external magnetic field--- correspond to a quark
effective mass $M=350$~MeV and a quark-antiquark condensate $\langle\bar ff
\rangle = (- 243.3\, \mbox{MeV})^3$. This parametrization, denoted as S350,
reproduces adequately the empirical values of $m_{\pi^0}$ and $f_{\pi}$ in
vacuum, namely $m_{\pi^0}=135$~MeV and $f_{\pi} = 92.4$~MeV.

Besides the usual NJL model framework, we also consider the case in which
the coupling constant $G$ has not a fixed value but changes with the
magnetic field. In fact, it is seen that NJL-like models usually fail to
describe the inverse magnetic catalysis effect observed in lattice QCD at
finite temperature (an interesting exception is the case of models which
include nonlocal interactions~\cite{Pagura:2016pwr,GomezDumm:2017iex}). To
cure this unwanted behavior, it is possible to assume that the strong
coupling $G$ depends on the magnetic field. This can be understood as an
effective way of incorporating the sea effect produced by the backreaction
of gluons to magnetized quarks loops. For definiteness, following
Ref.~\cite{Avancini:2016fgq} we consider a dependence $G(B)$ of the form
\begin{equation}
G(B) \ = \ G(0) \left[\,c_1+(1-c_1)\, e^{-c_2(eB)^2}\right] \ ,
\end{equation}
with $c_1 = 0.321$ and $c_2 = 1.31$~GeV$^{-2}$.

\begin{figure}[htb]
%%%%    \centering{}\includegraphics[width=0.48\textwidth]{massgcte}
%%%%    \centering{}\includegraphics[width=0.48\textwidth]{massgvar}
    \centering{}\includegraphics[width=0.85\textwidth]{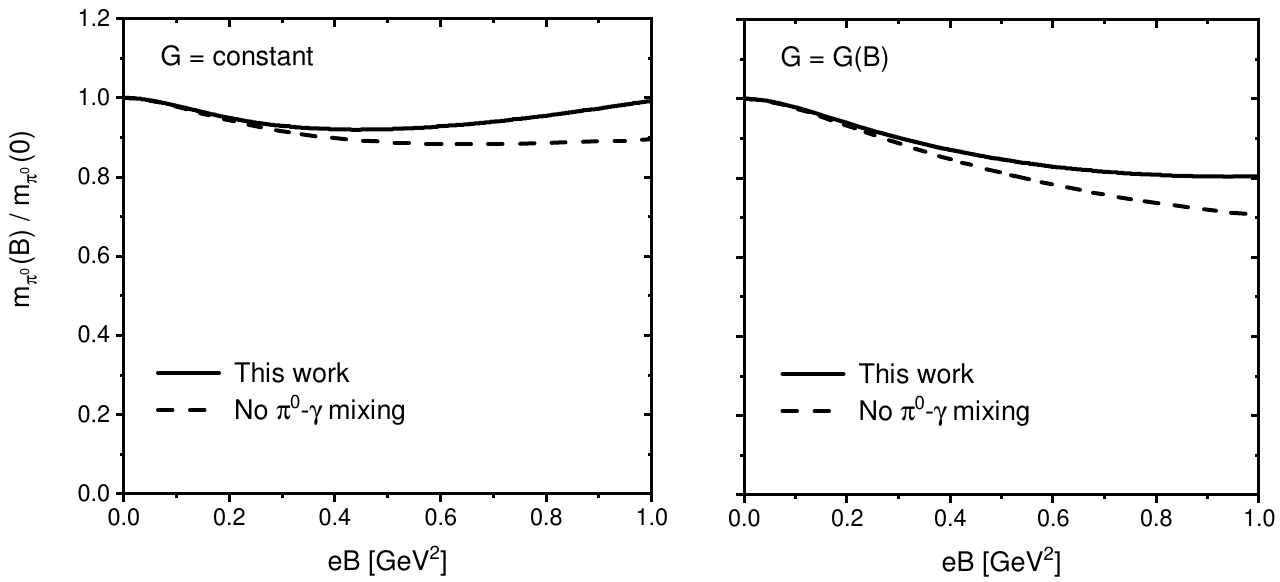}
\caption{$\pi^0$ mass as a function of the external magnetic
field (solid lines), for the cases $G=\;$constant and $G=G(B)$. Results for
the models with no $\pi^0$-$\gamma$ mixing (dashed lines) are shown for
comparison.}
\label{pimass}
\end{figure}

Our results for the $\pi^0$ mass, obtained from Eq.~(\ref{eqpimass}), are
shown in Fig.~\ref{pimass}. Left and right panels correspond to the models
with constant and $B$-dependent couplings $G$, respectively. In the case of
constant $G$, the behavior of the neutral pion mass with the magnetic field
is non-monotonic: while a minimum is shown at $eB\simeq 0.5$~GeV$^2$, the
value $m_{\pi^0}(0)$ is approximately recovered for $eB\simeq 1$~GeV$^2$. On
the other hand, for $G=G(B)$ the value of $m_{\pi^0}$ decreases steadily
with $B$ for the studied range of magnetic fields, reaching a value of about
$0.8\,m_{\pi^0}(0)$ for $eB\simeq 1$~GeV$^2$. By comparing with the results
obtained for no $\pi^0$-$\gamma$ mixing (dashed lines in both panels of
the figure), it is seen that in both cases the mixing leads to a relative
enhancement of the $\pi^0$ mass. This enhancement gets increased with the
magnetic field, reaching about 10 - 14\% for $eB\simeq 1$~GeV$^2$. It is
worth noticing that the effect is much less significant than the one
obtained in Ref.~\cite{Brauner:2017uiu}, where the mixing leads to a
relative enhancement of $\sim 90$\% in the $\pi^0$ mass for $eB\simeq
1$~GeV$^2$.

\begin{figure}[htb]
    \centering{}\includegraphics[width=0.5\textwidth]{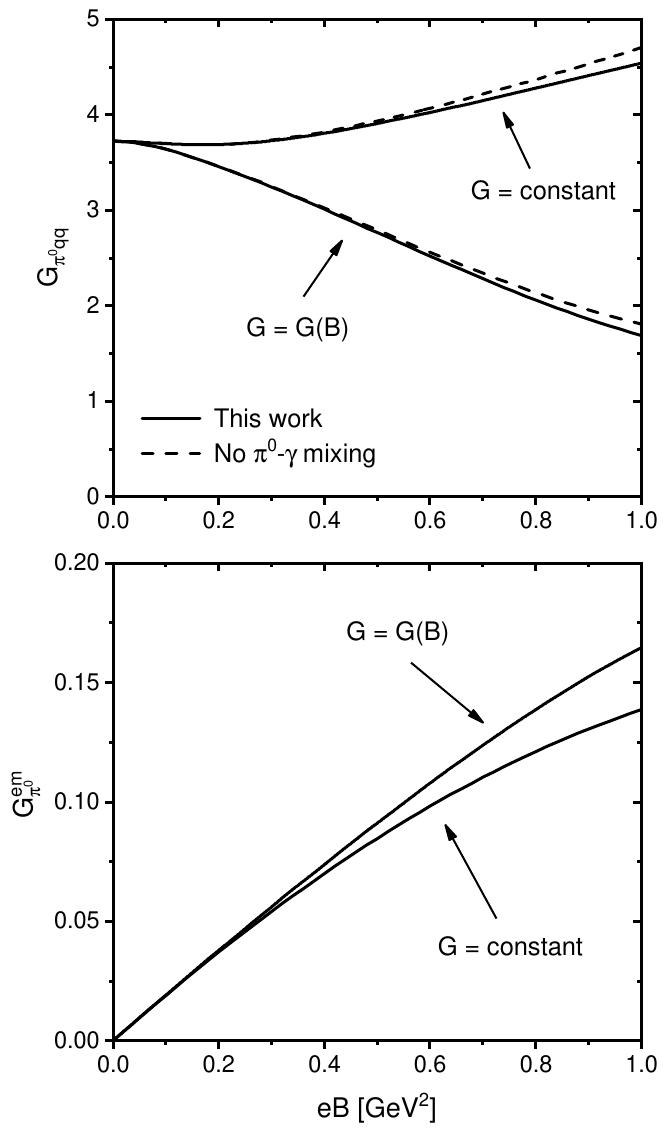}
\caption{$G_{\pi^0qq}$ (upper panel) and $G^{\rm em}_{\pi^0}$ (lower panel)
effective couplings as functions of the external magnetic field, for the
cases $G=\;$constant and $G=G(B)$. Values of $G_{\pi^0qq}$ for the models
with no $\pi^0$-$\gamma$ mixing (dashed lines) are shown for comparison.}
\label{gpiqqfig}
\end{figure}

In Fig.~\ref{gpiqqfig} we show the behavior of the effective
$\pi^0$-fermion couplings $G_{\pi^0qq}$ and $G^{\rm em}_{\pi^0}$ as
functions of the magnetic field. We include in the graphs the results
corresponding to the models with $G=\;$constant and $G=G(B)$. In addition,
for comparison we include the results for $G_{\pi^0qq}$ in the case of no
$\pi^0$-$\gamma$ mixing (dashed lines). The comparison between solid and
dashed lines in the upper panel show that there is no significant effect on
$G_{\pi^0qq}$ arising from the mixing. On the other hand, it is seen that
the behavior of $G_{\pi^0qq}$ for the case of constant $G$ is found to be
opposite to the one for $G=G(B)$, in agreement with the results obtained in
Ref.~\cite{Avancini:2016fgq}, where no $\pi^0$-$\gamma$ mixing was
considered. Regarding the effective coupling $G^{\rm em}_{\pi^0}$ (lower
panel of Fig.~\ref{gpiqqfig}), it is found that the behavior is similar for
$G=\;$constant and $G=G(B)$, reaching values of about $0.14$ and $0.16$,
respectively, for $eB=1$~GeV$^2$. It is observed that the values of the
coupling $G^{\rm em}_{\pi^0}$ ---which corresponds to the photon
contribution to the $\tilde \pi^0_R$-fermion vertex, and, therefore, is zero
if there is no $\pi^0$-$\gamma$ mixing--- are much smaller than those of
$G_{\pi^0qq}$ for the full considered range of values of the magnetic field.

\section{Summary and conclusions}

We have analyzed the mixing between the $\pi_0$ meson and photon fields in
the presence of a strong uniform magnetic field. Our calculations have been
carried out in the context of a two-flavor NJL model, considering both the
cases of a constant and a $B$-dependent four-quark effective coupling $G$.
The effect of the external magnetic field has been taken into account
consistently in all involved polarization functions.

As a main conclusion, our numerical analysis indicates that, contrary to
what has been stated in previous analyses, the effect of the
$\pi^0$-$\gamma$ mixing on the mass of the $\pi^0$ meson turns out to be
relatively small. For the considered range of values of $eB$ (up to
1~GeV$^2$), the correction with respect to the model with no mixing reaches
a maximum value of about 14\% for the case $G=G(B)$, and it is somewhat
lower for the case of constant $G$. Regarding the $\pi^0$-quark coupling
$G_{\pi qq}$, the effect is even smaller, reaching maximum values of about
6\% for the case $G=G(B)$.

It is worth mentioning that the $\pi^0$ field mixes only with one specific
photon polarization state. This can be understood from the fact that the
system is invariant under rotations around the axis defined by the
orientation of the magnetic field. For a pion at rest, the total spin
projection on that axis (say, axis 3) has to be a conserved quantity, and
the quantum number $S_{3}$ turns out to be a good one to characterize the
photon and pion states. As a consequence, it is seen that the pion can only
mix with the photon polarization state $\epsilon^{\mu}(2)$, defined by
Eq.~(\ref{polstates}).

\section*{Appendices}

\setcounter{section}{0}
\global\long\def\thesection{\Alph{section}}%
\global\long\def\theequation{\thesection.\arabic{equation}}%
\setcounter{equation}{0}
\global\long\def\thesubsection{\thesection.\arabic{subsection}}%

\section{Inverse propagator matrix in the pion rest frame}

\label{app1}

The integrals $C_{\pi\pi}^{f,{\rm mag}}$, $C^f_{\pi\gamma}$,
$C_{\gamma\gamma,i}^f$ and $C_{\gamma\gamma,0}^{f,{\rm mag}}$ appearing in
the inverse propagator matrix turn out to be simplified in the frame where
$q_\parallel^2=m_\pi^2$, $q_\perp^2=0$. From Eq.~(\ref{cppfull}) we find
%, (\ref{cpigam}), (\ref{c12}) and (\ref{c1mag}), respectively,
\begin{eqnarray}
\hspace{-0.5cm} C_{\pi\pi}^{f,{\rm mag}}(m_\pi^2,0) &=&
\ \frac{N_{c}}{8\pi^{2}}\int_{0}^{\infty}dz\,\int_{-1}^{1}d\beta\
e^{-z\,\omega_0(m_\pi^2)}\;
\nonumber\\
&& \times \ \bigg[
\left(M^{2}+\frac{1}{z}+\frac{1-\beta^{2}}{4}\;m_\pi^{2}\right) \left(
\frac{B_{f}}{\tanh(zB_{f})}
-\frac{1}{z}\right)+\,\frac{B_{f}^{2}}{\sinh^{2}(zB_{f})} -\frac{1}{z^2}
 \bigg]\ .
\end{eqnarray}
Upon integration by parts, this quantity can alternatively be written as
\begin{equation}
C_{\pi\pi}^{f,{\rm mag}}(m_\pi^2,0) \ = \  N_{c}\left[I_{1f}^{\rm mag} - m_\pi^{2}\,
I^{\rm mag}_{2f}(m_\pi^2)\right]\ ,
\end{equation}
where $I_{1f}^{\rm mag}$ is given by Eq.~(\ref{i1}), while the function
$I^{\rm mag}_{2f}(q^2)$ reads
\begin{eqnarray}
I^{\rm mag}_{2f}(q^2) =  - \dfrac{1}{16\pi^2} \int_{-1}^1 d\beta
\int_0^\infty dz  \:  e^{ - z \omega_0(q^2)/B_f}  \left(\coth z - \dfrac{1}{z} \right)\ .
\label{i2uno}
\end{eqnarray}
If, as expected, quark effective masses $M$ are large enough to satisfy
$m_\pi < 2M$, the above integral is convergent. The integration over $z$ can
be carried out analytically, leading to
\begin{equation}
I^{\rm mag}_{2f}(m_\pi^2) \ = \ \dfrac{1}{16\pi^2} \, \int_{-1}^{1} d\beta \left[
\psi(r_{\pi f}) - \ln \, r_{\pi f} + \dfrac{1}{2 \, r_{\pi f}}
\right] \ ,
\label{ImagAppB}
\end{equation}
where $\psi(x)$ is the digamma function, and we have defined
\begin{equation}
r_{\pi f} \ = \ \frac{\omega_0(m_\pi^2)}{2B_f}
\ = \ \frac{M^2-(1-\beta^2)\,m_\pi^2/4}{2B_f}\ .
\end{equation}

In turn, for the $\pi$-$\gamma$ mixing entry we have from Eq.~(\ref{cpigam})
\begin{eqnarray}
C^f_{\pi\gamma}(m_\pi^2,0) \ = \ \frac{1}{8\pi^{2}}\, N_{c}\, s_{f} \, M\,
\kappa_f \,Q_f\, B_{f} \int_{-1}^{1}d\beta\ \int_0^{\infty}dz\;
e^{-z\,\omega_0(m_\pi^{2})}\ .
\end{eqnarray}
These integrals can be performed analytically. Assuming $m_\pi < 2 M$ we
get
\begin{equation}
C^f_{\pi\gamma}(m_\pi^2,0)\ = \ \frac{N_{c}\, s_{f} \, M\,
\kappa_f \,Q_f\, B_{f}}{\pi^{2} \,m_\pi^2 \,\sqrt{4 M^2/m_\pi^2 -1}}
\,\arctan\left(\frac{1}{\sqrt{4 M^2/m_\pi^2 -1}}\right)\ .
\end{equation}

Finally, for the $\gamma$-$\gamma$ polarization functions we have to
consider the integrals in Eqs.~(\ref{c12}-\ref{c1reg}). Changing the
integration variable $z$ to $\tau = z B_f$ and using the above definition
for $r_{\pi f}$ we have
\begin{eqnarray}
C_{\gamma\gamma,0}^{f,{\rm mag}}(m_\pi^2,0) &=& - \frac{N_c\, Q_f^2}{16 \pi^2} \,
\int_{-1}^{1}d\beta\,\int_{0}^{\infty}d\tau
\ e^{-2\tau\,r_{\pi f}}\, \gamma_0(\tau,\beta) \ , \label{c1magpi} \\
C_{\gamma\gamma,i}^f(m_\pi^2,0) &=& - \frac{N_c\, Q_f^2}{16 \pi^2} \,
\int_{-1}^{1} d\beta\,\int_{0}^{\infty} d\tau
\ e^{-2\tau\,r_{\pi f}} \Big[\gamma_i(\tau,\beta)\, -
\, \gamma_0(\tau,\beta)\Big]\ , \quad \ i=1,2\ , \label{c12pi}
\end{eqnarray}
where
\begin{eqnarray}
\gamma_0(\tau,\beta) & = &\frac{ \cosh(\tau\beta) - \beta \coth \tau\,
\sinh(\tau\beta)}{\sinh\tau} \, - \, \frac{1-\beta^2}{\tau}\ , \nonumber
\\[2mm]
\gamma_1(\tau,\beta) & = & (1 - \beta^2)\left( \coth\tau\, -
\,\frac{1}{\tau}\right)
\ , \nonumber \\[2mm]
\gamma_2(\tau,\beta) & = & \frac{ 2 \big[\!\cosh\tau -
\cosh(\tau\beta)\big]}{\sinh^3\!\tau} \, - \, \frac{1-\beta^2}{\tau}\ .
\end{eqnarray}
The integrals over $\tau$ can be carried out taking into account the
relations
\begin{eqnarray}
\int_0^\infty d\tau\ e^{-2\lambda\tau} \left(\frac{\cosh (\tau\beta)}{\sinh\tau} \, -\,
\frac{1}{\tau}\right)
&=&
\ln \lambda\, -\, \frac{1}{2} \left[ \psi\left(\frac{2\lambda + \beta+1 }{2}\right)
+ \psi\left(\frac{2\lambda - \beta+1 }{2}\right) \right] \ ,
\nonumber \\[2mm]
\int_0^\infty d\tau\ e^{-2\lambda\tau} \; \frac{\sinh (\tau\beta)}{\sinh\tau}
&=& \frac{1}{2} \left[ \psi\left(\frac{2\lambda + \beta+1 }{2}\right)
+ \psi\left(\frac{2\lambda - \beta+1 }{2}\right) \right] \ ,
\end{eqnarray}
where we have used the recurrence relation $\psi(x+1) =
\psi(x)+1/x$. We obtain
\begin{eqnarray}
\int_{-1}^1 d\beta \int_0^\infty d\tau \ e^{-2\tau\, r_{\pi f}}\, \gamma_0(\tau,\beta) &=&
\int_{-1}^1 d\beta\, \bigg[ - \beta^2 + (1-\beta^2) \ln r_{\pi f} \nonumber \\
& & \hspace{1.cm} +\,
(\beta^2 + 2 \beta\, r_{\pi f} -1)\ \psi\left(r_{\pi f} +
\frac{\beta+1}{2}\right)\bigg]\ ,
\nonumber \\[2mm]
\int_{-1}^1 d\beta \int_0^\infty d\tau \ e^{-2\tau\, r_{\pi f}}\, \gamma_1(\tau,\beta) &=&
\int_{-1}^1 d\beta\;(1-\beta^2) \bigg[ \ln r_{\pi f} - \psi(r_{\pi f}) -\,
\frac{1}{2r_{\pi f}}\bigg]\ ,
\nonumber \\[2mm]
\int_{-1}^1 d\beta \int_0^\infty d\tau \ e^{-2\tau\, r_{\pi f}}\, \gamma_2(\tau,\beta) &=&
\int_{-1}^1 d\beta\, \bigg[ -2 +(1- \beta^2) \,\left(\frac{3}{2}\, +\, \ln r_{\pi f}\right)
- 2\,r_{\pi f} - 4\,r_{\pi f}^2\,\psi(r_{\pi f})\nonumber \\
& & \hspace{1.cm}
 +\, \big[(\beta + 2\, r_{\pi f})^2 -1\big]\ \psi\left(r_{\pi f} +
\frac{\beta+1}{2}\right)\bigg]\ .
\end{eqnarray}

Now, from Eqs.~(\ref{cmagreg}) and (\ref{g2g2}) it is seen that the matrix
element $\gmatrix_{\gamma_2\gamma_2}$ needed to determine the $\pi^0$ mass
is given by
\begin{equation}
{\gmatrix}_{\gamma_2\gamma_2} \ = \
- m_\pi^2  + \sum_f \Big[
C_{\gamma\gamma,0}^{f,0,{\rm reg}}(m_\pi^2)\, + \,
C_{\gamma\gamma,0}^{f,{\rm mag}}(m_\pi^2,0)\, + \,
C_{\gamma\gamma,1}^f(m_\pi^2,0)\Big]\, m_\pi^2 \ .
\end{equation}
According to the above equations,
% Eqs.~(\ref{c1magpi}) and (\ref{c12pi}),
the sum of the ``magnetic'' contributions is given by
\begin{eqnarray}
C_{\gamma\gamma,0}^{f,{\rm mag}}(m_\pi^2,0)\, + \,
C_{\gamma\gamma,1}^f(m_\pi^2,0) &=& - \frac{N_c Q_f^2}{16\pi^2} \int_0^\infty d\tau
\ e^{-2\tau\,r_{\pi f}} \,\gamma_1(\tau,\beta)
\nonumber \\
&=& - \frac{N_c Q_f^2}{16\pi^2} \int_{-1}^1 d\beta \ (1-\beta^2) \left[ \ln r_{\pi f} - \psi(r_{\pi f})
- \frac{1}{2 r_{\pi f}} \right]\ .
\end{eqnarray}
Adding now the regularized ``$B=0$'' contribution $C_{\gamma\gamma,0}^{f,0,{\rm reg}}(m_\pi^2)$ given by
Eq.~(\ref{c0reg}), the log terms can be trivially integrated, and we end up
with
\begin{equation}
\gmatrix_{\gamma_2\gamma_2} \ = \ -\, m_\pi^2  + \, \frac{N_c\,m_\pi^2}{16 \pi^2}
\sum_f\,Q_f^2
\left\{ - \frac{4}{3} \, \ln \left(\frac{M^2}{2B_f}\right)
+ \int_{-1}^1 d\beta\ (1-\beta^2) \left[ \psi (r_{\pi f}) +
\frac{1}{2 r_{\pi f}}\right] \right\}\ .
\label{gg2g2_quarks}
\end{equation}

The above expression is valid as long as the condition $m_\pi<2M$ is
fulfilled, which is expected to hold for quark loops. Now, as stated in the
main text, to determine the pion mass one has to take into account all
fermion loop contributions to $\gmatrix_{\gamma_2\gamma_2}$, which means to
add QED lepton loops. These are similar to those included in the sum in
Eq.~(\ref{gg2g2_quarks}): one has just to replace $N_c\to 1$, $Q_f^2\to e^2$
and $M\to m_l$, where $l=e$, $\mu$ (the contribution of the $\tau$ lepton
can be safely neglected). In the case of the electron loop, the $\pi^0$ mass
lies above the threshold of $e^+e^-$ production. Therefore, the contribution
of the loop to $\gmatrix_{\gamma_2\gamma_2}$ will have an absorptive part.
As it is usually done, we addressed this problem by carrying out an
analytical extension of the integral in Eq.~(\ref{gg2g2_quarks}). We define
the function
\begin{equation}
F(q^2) \ = \ \int_{-1}^1 d\beta\ (1-\beta^2) \left[ \psi (r(q^2)) +
\frac{1}{2 r(q^2)}\right] \ ,
\end{equation}
where $r(q^2) = [4 m_e^2-(1-\beta^2)\,q^2]/(8B_e)$, with $B_e = |eB|$.
Performing the analytic extension to the region $q^2>4m_e^2$ one has (see
e.g.~Ref.~\cite{Carlomagno:2022inu})
\begin{eqnarray}
F(q^2) & = & \int_{-1}^1 d\beta\ (1-\beta^2) \;\psi \Big( r(q^2) + N + 1 \Big)
\, +\, \frac{8 B_e}{q^2} \sum_{n=0}^N \frac{2-\delta_{n0}}{\rho_n}
\Big[\,\rho_n+ (1-\rho_n^2) \arctanh \rho_n\Big]
\nonumber \\
&&  - \,i \, \frac{4 \pi B_e}{q^2}
\sum_{n=0}^N \frac{2-\delta_{n0}}{\rho_n}\, (1-\rho_n^2)\ ,
\end{eqnarray}
where we have introduced the definitions
\begin{equation}
\rho_n = \sqrt{1\, -\, \frac{4(m_e^2  + 2 n B_e)}{q^2}}\ ,
\qquad\qquad N = \mathrm{Floor} \left( \dfrac{q^2- 4 m_e^2}{8 B_e}\right)\ .
\end{equation}

\end{document}